\input{aipcheck.tex}

\newcommand\sps{\space\space\space\space}
  \def\selectedoptions{final}

\documentclass[
   \selectedoptions
  ]
  {aipproc}

\usepackage{amssymb}
\usepackage{amsmath}

  \def\selectedlayoutstyle{6x9}
\layoutstyle\selectedlayoutstyle

\SetInternalRegister\hbadness{8000} 

\newcommand\doingARLO[2][]{%
  \ifx\mmref\undefined #1\else #2\fi
}
  
\def\be{\begin{equation}}
\def\ee{\end{equation}}
\def\bea{\begin{eqnarray}}
\def\eea{\end{eqnarray}}

\begin{document}

\title 
      [Distinguishing between $w < -1$]
      {Distinguishing between $w < -1$ Dark Energy Models}


\author{Houri Ziaeepour}{
  address={Mullard Space Science Laboratory, Holmbury St. Mary, Dorking, 
Surrey, RH5 6NT, UK.},
  email={hz@mssl.ucl.ac.uk},
}

\copyrightyear  {2007}

\begin{abstract}
Recent data and new data analysis methods show that most probably the 
parameter $w$ in the equation of state of the dark energy is smaller 
than $-1$ at low redshifts. We briefly review some of the models with such a 
property and without violating null energy condition. We investigate the 
difference between the observables and predictions of these models, and how 
they can be explored to single out or constrain the origin of dark energy 
and its properties.
\end{abstract}

\date{30 Aug. 2007}

\maketitle


Understanding the nature of dark energy is one of the biggest challenges in present 
cosmology and particle physics. In order to achieve this goal, the 
measurement of the cosmological evolution of the dark energy, parameters of 
the candidate models, and their origin and relation with other contents of the 
Universe are of extreme importance. This also means that the way we model the 
data and extract parameters affects our interpretation of what is the dark 
energy and how it evolves. 

At present all the determination of dark energy parameters is based on the 
simplest extension of the LCDM. The energy content of the Universe is usually 
considered to be composed of 3 components: cold visible and dark matter, 
hot matter, and a dark energy component with a perfect fluid equation of state:
\be
\frac {H^2 (z)}{H_0^2} = \frac{\rho (z)}{\rho_0} = \Omega_{m} (1+z)^3 + 
\Omega_{hot} (1+z)^4 + \Omega_{de} (1+z)^{3\gamma (z)}
\label {rhoclassic}
\ee
When $\gamma (z)$ does not depend on redshift, $\gamma = w + 1$, and 
$w \equiv P/\rho$. We note that in this definition no interaction between various 
components is included. Although observations of the CMB anisotropy shows 
that the non-gravitational interaction between the hot matter - mainly CMB - 
and dark and baryonic matter is very small, the constraints on the 
non-gravitational interaction between dark components are not very 
strong\cite{deaniso}. 

\begin{figure}
\includegraphics[height=6.5cm,angle=-90]{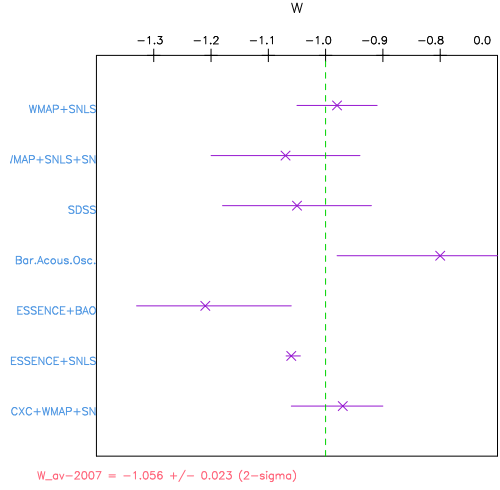}
\caption{Recent estimations of $w$ and their weighted average. \label{fig:w07}}
\end{figure}

Using \ref{rhoclassic} for fitting data from CMB, LSS, and supernovae, the 
estimations of recent measurements are summarized in Fig.\ref{fig:w07}. Many 
of these estimations relay on multiple type of data to remove the degeneracy 
between cosmological parameters.  

Therefore in this sense not all of them are 
independent. Nonetheless, if we assume that the difference in their estimation 
is statistical, 
the best estimation of $w$ would be the weighted average of all the measurements. 
What one obtains in this way is $w = -1.056 \pm 0.023$ at $2\sigma$ level. 
This value as well as the individual measurements are all very close to the 
critical value $w = -1$ (a cosmological constant). Nevertheless, the best 
estimation and most of the measurements are consistent with $w \lesssim -1$. 
In the most simplistic view of the data we can say that the deviation from a 
cosmological constant is due to the errors. However, considering the best 
estimation mentioned above, even at $2\sigma$ level $w < -1$. This can be 
the evidence of a much richer physics behind the dark energy than just a putative 
cosmological constant.

Evidently just one number is not enough to decide about the nature of such 
a complex entity as the dark energy. There have been a number of attempts to 
estimate the redshift evolution of $w$. Although present data is not ideal 
for this purpose, the minimal conclusion is that it does not significantly  
vary up to $z \sim 1$\cite{dezvar}. Here we want also to mention very briefly 
the issue 
of the techniques usually employed to determine the equation of state of the dark 
energy from data. In most data sets one has to fit a large number of parameters 
including $w$ or its redshift expansion parameters together. Although this 
method is acceptable for quantities that their exact value is not crucial for 
their physics such as $\Omega_{dm}$ - at least not at our present level of 
knowledge e- the degeneracy in fit results can be very important for parameters 
close to a critical value such as $w$. A method based on geometrical 
properties of $d\rho/dz$ is suggested ~\cite{hourinonpara} which in one 
hand permits to directly measure the sign of $w$ and its redshift dependence. 
On the other hand, up to certain limits its results are less affected by the 
uncertainty in prior parameters such as $H_0$ and $\Omega_{m}$. Application of 
this method to the publicly available SN type I data shows that $w < -1$ at least 
up to $z \sim 0.5$ and its variation, if exists, is very small.

The main argument - or rather we can call it ``fear'' - against $w < -1$ is that it 
violates the weak energy condition i.e. $P_{de}+\rho_{de} \geqslant 0$. However, 
we should remember that this condition is written for a perfect fluid without 
interaction. If the interaction between various components, in particular with 
the dark energy, is taken into account the form of null energy condition would be 
much more complex and eventually $w$ can be less than $-1$ without violation 
of any fundamental law of physics. In this case the evolution of total density 
can be written as:
\be
\frac{\rho (z)}{\rho_0} = \Omega_{m} (1+z)^3 + \Omega_{hot} (1+z)^4 + 
\Omega_{de} (1+z)^{3\gamma (z)} + g^2 f (\Omega_{m}, \Omega_{hot}, 
\Omega_{de}, z) \label{rhoint}
\ee
where $f$ is an unknown function and depends on the dark sector model. It is 
possible to show that for a number of models in which dark matter and dark 
energy interact with each other or are not an ideal fluid, {\it the effective} 
$w$ when \ref{rhoclassic} is used to analyze the data in place of \ref{rhoint}, 
is smaller than $-1$. Here we mention a few examples of these models. The 
simplest example is a decaying dark matter with cosmological constant as dark 
energy\cite{houriustate}. The next one is a decaying dark matter with a very 
small branching ratio to a light, axion-like scalar\cite{houridmquin,houridmquin1}. 
The condensation of this field plays the role of a quintessence field with an 
evolution very similar to a cosmological constant from very early times after 
formation of the meta-stable dark matter. Finally the last example we mention 
here is a stable dark matter interacting with a quintessence field as dark 
energy\cite{quinint}. Many other examples can be found, but at least some of 
them imply a violation of equivalence principal or cosmological variation of 
particle masses and couplings which are strongly constrained by the non-observation 
of density dependent effects of dark energy at cosmological distances. 

How can we distinguish between these models ?  
This task is specially more difficult in the situation where dark energy does 
not strongly vary with redshift. But even if it has some variation, it is 
very difficult to conclude the nature of the underlaying model just from a 
couple of parameters that determine its variation. We need additional 
observables more closely related to the field theoretical aspects of the models. 
At present level of our 
knowledge about the physics beyond the Standard Model, it is very difficult 
to place the dark energy field - quintessence scalar - in the zoo of particles. 
Nonetheless, it is possible to guess some of the possible observables. For 
instance, if dark matter has a non-gravitational interaction with dark energy 
condensate, we expect that {\it dark energy particles} are released from the 
condensate. As they are expected to be very light, they should make a hot dark 
background of non-SM nature. This process is very similar to the scattering 
from a Bose-Einstein condensate\cite{bsescatter}. The cosmological density of 
this hot component and its evolution depends on the type and strength of the 
interaction. These are unknown, but from constraints on the clustering of the 
dark energy we expect that the coupling must be very small, and therefore the 
density of the corresponding hot matter should be small too - probably less 
than CMB, nonetheless the observation of such a component can significantly 
help to understand the origin of dark energy. On the other hand, we should 
also expect some anisotropy in the dark energy at large that may be observable 
specially in the high precision CMB data.

If the dark energy is a cosmological constant the situation is more 
ambiguous because we don't yet have a generally acceptable definition for a 
vacuum energy. If its origin is some physics at very high energy 
scale - Planck or superstring scale - and fixed at very early stage in the 
evolution of the Universe, the dark energy would be completely static, 
isotropic, and have no effect other than gravity up to any measurable 
redshift. In this 
case an {\it effective} $w < -1$ would be the signature of a non-gravitational 
interaction in dark matter sector, for instance slow decay of dark matter. 
If at least part of the remnants of the decay are SM particles, we must be able 
to observe them as an additional component in cosmic rays. Their energy range 
depends on the mass of the decaying dark matter particles and can be 
$E \gtrsim 10^{13}$~GeV or $E \lesssim 100$~GeV. Moreover, it must somehow 
correlate with the large structures. By contrast, if the remnants 
are all dark, then they can be only detected through their effect on 
the structure formation. But this can be very small and very difficult 
to observe.

The last case we consider here is when the dark energy is the result of the 
condensation of a light scalar produced during the decay of a meta-stable dark 
matter. It has been shown that the equation of state of the condensate is 
very similar to a cosmological constant~\cite {houridmquin}. This model can 
produce a light hot dark matter - the scalar particles - an excess of 
visible cosmic rays, and/or dark remnants. The anisotropy of the dark energy 
would be very small but worth to be searched for. 

\begin{figure}
\includegraphics[height=3.5cm]{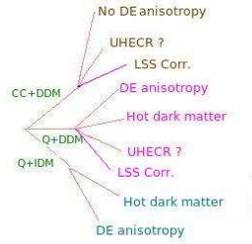}
\caption {Dark energy models with effective $w < -1$ and their observables. 
\label {fig:diag}}
\end {figure}

Fig.\ref{fig:diag} summarizes the possible origins of a dark energy with 
effective $w < -1$ considered here and their observables.

Evidently, the most important question is how we can observe the tiny 
and weak interacting observables explained above. In one hand, this needs 
drastic improvement in the precision of measurements. On the other hand, more 
precise observations of cosmological effects means more sensitivity to the 
foreground phenomena that can mislead interpretation of observations. 
Therefore, the first step in this direction is a better understanding of 
the foreground. On another front, the discovery of physics beyond the Standard 
Model by LHC can be very important for clarifying which direction(s), both 
theoretical and experimental, we should investigate more closely to find the 
origin of the dark energy.

In conclusion, we have discussed some of the possible models for dark energy 
with effective $w < -1$ and without violation of the null energy condition. At 
present these types of dark energy are preferred by the data. Without 
considering any explicit implementation of these models we investigated the 
difference between their observables and how this can help to pin-down 
the underlaying model.


\doingARLO[\bibliographystyle{aipproc}]
          {\ifthenelse{\equal{\AIPcitestyleselect}{num}}
             {\bibliographystyle{arlonum}}
             {\bibliographystyle{arlobib}}
          }
\bibliography{proceedingsaip}

\end{document}